\documentclass[aps,prd,preprintnumbers,superscriptaddress,nofootinbib,showpacs,twocolumn]{revtex4}%
\usepackage[]{graphicx}
\usepackage{bm,latexsym,amsmath,amssymb,amsfonts,mathrsfs}
\newcommand*{\D}{{\rm d}}
\newcommand*{\mpl}{M_{\rm Pl}}
\newcommand*{\msun}{M_{\odot}}

\begin{document}

\title{Relativistic stars in degenerate higher-order scalar-tensor theories
after GW170817}

\author{Tsutomu~Kobayashi}
\email[Email: ]{tsutomu@rikkyo.ac.jp}
\affiliation{Department of Physics, Rikkyo University, Toshima, Tokyo 171-8501, Japan}

\author{Takashi~Hiramatsu}
\email[Email: ]{hiramatz@rikkyo.ac.jp}
\affiliation{Department of Physics, Rikkyo University, Toshima, Tokyo 171-8501, Japan}

\begin{abstract}
We study relativistic stars in degenerate higher-order
scalar-tensor theories that evade the constraint
on the speed of gravitational waves imposed by GW170817.
It is shown that
the exterior metric is given by the usual Schwarzschild solution
if the lower order Horndeski terms are ignored in the Lagrangian and
a shift symmetry is assumed.
However, this class of theories exhibits partial breaking of Vainshtein
screening in the stellar interior and thus modifies the structure of a star.
Employing a simple concrete model, we show that for high-density stars
the mass-radius relation is altered significantly even if
the parameters are chosen so that only a tiny correction is expected
in the Newtonian regime. We also find that, depending on the parameters,
there is a maximum central density above which
solutions cease to exist.
\end{abstract}

\preprint{RUP-18-10}
\maketitle

\section{Introduction}

The nearly simultaneous detection of
gravitational waves GW170817
and the $\gamma$-ray burst GRB 170817A~\cite{TheLIGOScientific:2017qsa,%
Monitor:2017mdv,GBM:2017lvd}
places a very tight constraint on the
speed of gravitational waves, $c_{\rm GW}$.
The deviation of $c_{\rm GW}$ from the speed of light
is less than 1 part in $10^{15}$.
This can be translated to constraints on
modified gravity such as scalar-tensor theories,
vector-tensor theories, massive gravity,
and Ho\v{r}ava gravity~\cite{Creminelli:2017sry,%
Sakstein:2017xjx,Ezquiaga:2017ekz,Baker:2017hug,Gumrukcuoglu:2017ijh,%
Oost:2018tcv}.
In particular, in the context of the Horndeski theory
(the most general scalar-tensor theory
having second-order equations of motion\footnote{The original
form of the Horndeski action is different from its modern expression
obtained by extending the Galileon theory~\cite{Deffayet:2011gz}.
The equivalence of the two apparently different theories
was shown in~\cite{Kobayashi:2011nu}.})~\cite{Horndeski:1974wa},
two of the four free functions in the action are strongly constrained,
leaving only the simple, traditional form of
nonminimal coupling of the scalar degree of freedom to
the Ricci scalar, {\em i.e.,} the ``$f(\phi){\cal R}$''-type coupling.

However, it has been pointed out that
there still remains an interesting, nontrivial class of scalar-tensor theories
beyond Horndeski
that can evade the gravitational wave constraint
as well as solar-system tests~\cite{Creminelli:2017sry,Ezquiaga:2017ekz,Sakstein:2017xjx,%
Crisostomi:2017lbg,Langlois:2017dyl,Babichev:2017lmw,%
Dima:2017pwp,Crisostomi:2017pjs,Bartolo:2017ibw}.
Such theories have higher-order equations of motion as they are
more general than the Horndeski theory,
but the system is degenerate and hence
is free from the dangerous extra degree of freedom that
causes Ostrogradski instability.
Earlier examples of degenerate higher-order scalar-tensor (DHOST) theory
are found in~\cite{Zumalacarregui:2013pma,Gleyzes:2014dya},
and the degeneracy conditions are systematically
studied and classified
at quadratic order in second derivatives of the scalar field
in~\cite{Langlois:2015cwa,Crisostomi:2016czh}
and at cubic order in~\cite{BenAchour:2016fzp}.
Degenerate theories can also be generated from
nondegenerate ones via noninvertible disformal
transformation~\cite{Takahashi:2017pje,Langlois:2018jdg}.

One of the most interesting phenomenologies of DHOST theories is
efficient Vainshtein screening outside matter sources
and its partial breaking in the inside~\cite{Kobayashi:2014ida}.
The partial breaking of screening modifies,
for instance, stellar structure~\cite{Koyama:2015oma,Saito:2015fza}.
This fact was used to test DHOST theories,
or, more specifically, the Gleyzes-Langlois-Piazza-Vernizzi (GLPV)
subclass~\cite{Gleyzes:2014dya},
against astrophysical observations~\cite{Sakstein:2015zoa,%
Sakstein:2015aac,Jain:2015edg,Sakstein:2016ggl,Sakstein:2016lyj,Salzano:2017qac}.
Going beyond the weak-field approximation,
relativistic stars in the GLPV theory have been studied
in~\cite{Babichev:2016jom,Sakstein:2016oel}.

In this paper, we consider relativistic stars in
DHOST theories that are more general than the GLPV theory
but evade the constraint on the speed of gravitational
waves~\cite{Creminelli:2017sry,Sakstein:2017xjx,Ezquiaga:2017ekz}.
So far, this class of theories have been investigated
by employing the weak-field
approximation~\cite{Crisostomi:2017lbg,Langlois:2017dyl,Dima:2017pwp}
and in a cosmological context~\cite{Crisostomi:2017pjs}.
Very recently, compact objects including relativistic stars
in the GLPV theory with $c_{\rm GW}=1$
have been explored in Ref.~\cite{Chagoya:2018lmv}.

This paper is organized as follows.
In the next section, we introduce the DHOST theories
with $c_{\rm GW}=1$ and
derive the basic equations describing a spherically symmetric
relativistic star. To check the consistency with the previous results,
we linearize the equations and see the gravitational potential
in the weak-field approximation in Sec.~III.
Then, in Sec.~IV,  we give boundary conditions
imposed at the stellar center and in the exterior region.
Our numerical results are presented in Sec.~V.
We draw our conclusions in Sec.~VI.
Since some of the equations are quite messy,
their explicit expression is shown in Appendix~A.

\section{Field equations}

The action of the quadratic DHOST theory we study is given by
\begin{align}
S=\int \D^4x\sqrt{-g}\left[f(X){\cal R}+\sum_{I=1}^5{\cal L}_I+{\cal L}_{\rm m}\right],
\label{eq:Lagrangian}
\end{align}
where ${\cal R}$ is the Ricci scalar, $X:=\phi_\mu \phi^\mu$, and
\begin{align}
{\cal L}_1&:=A_1(X)\phi_{\mu\nu}\phi^{\mu\nu},
\\
{\cal L}_2&:=A_2(X)(\Box\phi)^2,
\\
{\cal L}_3&:=A_3(X)\Box\phi\phi^\mu\phi_{\mu\nu}\phi^\nu,
\\
{\cal L}_4&:=A_4(X)\phi^\mu\phi_{\mu\rho}\phi^{\rho \nu}\phi_\nu,
\\
{\cal L}_5&:=A_5(X)(\phi^\mu\phi_{\mu\nu}\phi^\nu)^2,
\end{align}
with $\phi_\mu=\nabla_\mu\phi$
and $\phi_{\mu\nu}=\nabla_\mu\nabla_\nu\phi$.
The functions $A_I(X)$ must be subject to certain conditions
in order for the theory to be degenerate and satisfy $c_{\rm GW}=1$,
as explained shortly.
Here, shift symmetry is assumed and
the other possible terms
of the form $G_2(X)$ and $G_3(X)\Box\phi$ are omitted.
In particular, we do not include the usual kinetic term $-X/2$
in this paper.\footnote{The inclusion of the usual kinetic term $-X/2$
gives rise to an interesting branch of asymptotically locally flat solutions
with a deficit angle, as has been recently explored in Ref.~\cite{Chagoya:2018lmv}.
We leave this possibility for future research.
The appearance of a deficit angle in the GLPV theory was
pointed out earlier in Refs.~\cite{DeFelice:2015sya,Kase:2015gxi}.
As we will see, our ansatz for the scalar field
is different from that assumed in Refs.~\cite{DeFelice:2015sya,Kase:2015gxi}
(see Eq.~(\ref{eq:vt}) below).
Probably this is the reason why we can obtain spherically symmetric solutions
that are regular at the center.}
These assumptions are nothing to do with
the degeneracy conditions and the $c_{\rm GW}=1$ constraint
to be imposed below.
However,
with this simplification one can concentrate on
the effect of Vainshtein breaking.

Note in passing that the DHOST theories
are equivalent to the Horndeski theory with disformally coupled
matter. Therefore, the setup we are considering
is in some sense similar to that explored in Ref.~\cite{Minamitsuji:2016hkk}.
However, the crucial difference is that, in contrast to
Ref.~\cite{Minamitsuji:2016hkk},
our theory corresponds to the case where the conformal
and disformal coupling functions depend on the first derivative of the
scalar field.

We require that
the speed of gravitational waves, $c_{\rm GW}$, is equal to
the speed of light~\cite{Monitor:2017mdv}.
In our theory
$c_{\rm GW}$ is given by $c_{\rm GW}^2=f/(f-XA_1)$~\cite{deRham:2016wji},
so that
\begin{align}
A_1=0.
\end{align}

The degeneracy conditions read
\begin{align}
A_2&=-A_1=0,
\\
A_4&=-\frac{1}{8f}\left(8A_3f-48f_X^2-8A_3f_XX+A_3^2X^2\right),\label{eq:degn1}
\\
A_5&=\frac{A_3}{2f}\left(4f_X+A_3X\right).\label{eq:degn2}
\end{align}
We thus have two free functions, $A_3$ and $f$, in the
quadratic DHOST sector with $c_{\rm GW}=1$.
This theory has been explored recently in
Refs.~\cite{Creminelli:2017sry,Sakstein:2017xjx,Ezquiaga:2017ekz,%
Baker:2017hug,Crisostomi:2017lbg,Langlois:2017dyl,Dima:2017pwp,Crisostomi:2017pjs}.
(See Ref.~\cite{Bartolo:2017ibw} for a different subclass
with $f=\;$const and $A_2=-A_1\neq 0$.)
Following Ref.~\cite{Crisostomi:2017pjs}, we introduce
\begin{align}
B_1:=\frac{X}{4f}(4f_X+XA_3),
\end{align}
and use this instead of $A_3$.
In the special case with $B_1=0=A_5$, the action
reduces to that of the GLPV theory.

We consider a static and spherically symmetric metric,
\begin{align}
\D s^2=-e^{\nu(r)}\D t^2+e^{\lambda(r)}\D r^2 +r^2\D\Omega^2.
\end{align}
The scalar field is taken to be
\begin{align}
\phi(t,r)=vt+\psi(r),\label{eq:vt}
\end{align}
where $v \,(\neq0)$ is a constant.
Even though $\phi$ is linearly dependent on
the time coordinate, it is consistent with the static spacetime
because the action~(\ref{eq:Lagrangian}) possess a shift symmetry,
$\phi\to\phi+c$, and $\phi$ without derivatives does not appear
in the field equations.
This ansatz was also used to obtain black hole solutions
in the Galileon and Horndeski theories in
Refs.~\cite{Babichev:2013cya,Kobayashi:2014eva,Babichev:2016rlq,%
Babichev:2016fbg,Babichev:2017guv}.

The field equations are given by
\begin{align}
{\cal E}_\mu^\nu&=T_\mu^\nu,\label{eq:grav-eq}
\\
\nabla_\mu J^\mu& =0,\label{eq:dJ=0}
\end{align}
where
${\cal E}_{\mu\nu}$ is obtained by varying the action with respect to the metric
and
$J^\mu$ is the shift current defined by
$\sqrt{-g}J^\mu=\delta S/\delta\phi_\mu$.
The energy-momentum tensor is of the form
\begin{align}
T_\mu^\nu={\rm diag}(-\rho, P,P,P).
\end{align}
The radial component of the conservation equations,
$\nabla_\mu T^{\mu}_\nu=0$, reads
\begin{align}
P'=-\frac{\nu'}{2}(\rho+P),\label{eq:fluid}
\end{align}
where ${}':=\D/\D r$.

With direct calculation we find that
\begin{align}
J^r\propto {\cal E}_{tr}.
\end{align}
Therefore, the gravitational field equation ${\cal E}_{tr}=0$
requires that $J^r$ vanishes.  Then,
the field equation for the scalar field~(\ref{eq:dJ=0})
is automatically satisfied.

To write Eq.~(\ref{eq:grav-eq}) and $J^r=0$ explicitly,
it is more convenient to use $X=-e^{-\nu}v^2+e^{-\lambda}\psi'^2$
instead of $\psi$.
In terms of $X$, we have
\begin{align}
{\cal E}_t^t&=b_1 \nu''+b_2X''+\widetilde{{\cal E}}_t(\nu,\nu',\lambda, \lambda', X, X'),
\label{eq:Ett}
\\
{\cal E}_r^r&=c_1 \nu''+c_2X''+\widetilde{{\cal E}}_r(\nu,\nu',\lambda, \lambda', X, X'),
\label{eq:Err}
\\
\psi' J^r&=c_1 \nu''+c_2X''+\widetilde{{\cal E}}_J(\nu,\nu',\lambda, \lambda', X, X'),
\label{eq:Jr}
\end{align}
where
\begin{align}
b_1&=2fB_1e^{-\lambda}\left(\frac{e^{-\nu}v^2}{X}\right),
\\
b_2&=b_1
\left[\frac{e^\nu}{v^2}+\frac{B_1(3X+4e^{-\nu}v^2)}{X^2} -\frac{4e^{-\nu}v^2f_X}{Xf}\right],
\\
c_1&=-2fB_1e^{-\lambda}\left(
\frac{e^{-\nu}v^2+X}{X}\right),
\\
c_2&=c_1
\left[ \frac{B_1(3X+4e^{-\nu}v^2)}{X^2} -\frac{4e^{-\nu}v^2f_X}{Xf} \right],
\end{align}
but the explicit expression of $\widetilde{{\cal E}}_t$,
$\widetilde{{\cal E}}_r$, and $\widetilde{{\cal E}}_J$ are messy.

We see that ${\cal E}_r^r$ and $\psi' J^r$
have the same coefficients $c_1$ and $c_2$.
Moreover, we find by an explicit computation
that ${\cal E}_r^r$ and $\psi' J^r$
are linearly dependent on $\lambda'$ and their coefficients are
also the same. Therefore, by taking the combination
${\cal E}_r^r-\psi' J^r$ one can remove $\nu''$, $X''$, and $\lambda'$.
Then, the field equation ${\cal E}_r^r-\psi'J^r=P$
can be solved for $\lambda$ to give
\begin{align}
e^\lambda = {\cal F}_\lambda (\nu,\nu',X,X',P),
\label{eq:elambda}
\end{align}
where
\begin{align}
{\cal F}_\lambda=\frac{2X+B_1rX'}{2X^3(2f+r^2P)}
 \bigl\{4e^{-\nu}v^2(fB_1-Xf_X)rX'&
\notag \\
 +Xf\left[3B_1rX'+2X(1+r\nu')\right]
\bigr\}.&
\end{align}

Using Eq.~(\ref{eq:elambda}), we can eliminate $\lambda$ and $\lambda'$
from Eqs.~(\ref{eq:Ett}) and~(\ref{eq:Jr}).
In doing so we replace $P'$ with $\nu'$, $\rho$, and $P$
by using Eq.~(\ref{eq:fluid}).
We then obtain
\begin{align}
\psi' J^r=k_1\nu''+k_2X''+{\cal J}_1(\nu,\nu',X,X',\rho,P)=0,
\label{eq:cJ}
\end{align}
where $k_{1,2}=k_{1,2}(\nu,\nu',X,X',P)$.
The field equation ${\cal E}_t^t+\rho=0$
can also be written in the form
\begin{align}
k_1\nu''+k_2X''+{\cal J}_2(\nu,\nu',X,X',\rho,P)=0.
\end{align}
Note that we have the same coefficients $k_1$ and $k_2$.
This is due to the degeneracy conditions.
We thus arrive at a first-order equation,
${\cal J}_1={\cal J}_2$, which can be solved for $X'$ as
\begin{align}
X'={\cal F}_1(\nu,X,\rho, P)\nu'+\frac{{\cal F}_2(\nu,X,\rho,P)}{r},
\label{eq:dX}
\end{align}
where ${\cal F}_1$ and ${\cal F}_2$ are complicated.
Their explicit form is presented in Appendix~\ref{App:f1f2f3}.
Finally, we use Eq.~(\ref{eq:dX})
to eliminate $X'$ and $X''$ from Eq.~(\ref{eq:cJ}).
This manipulation also removes $\nu''$, as it should be
because the theory is degenerate.
We thus arrive at
\begin{align}
\nu'={\cal F}_3(\nu,X,\rho,\rho',P),\label{eq:dnu}
\end{align}
where the explicit expression of ${\cal F}_3$ is extremely long
and is presented in Appendix~\ref{App:f1f2f3}.

We have thus obtained our basic equations
describing
the Tolman-Oppenheimer-Volkoff system
in DHOST theories.
Given the equation of state relating $\rho$ and $P$,
one can integrate Eqs.~(\ref{eq:fluid}), (\ref{eq:dX}), and~(\ref{eq:dnu})
to determine $P=P(r)$, $\nu=\nu(r)$, and $X=X(r)$.
Equation~(\ref{eq:elambda}) can then be used to obtain $\lambda=\lambda(r)$.

\section{Nonrelativistic, weak-field limit}

Since our procedure to obtain spherically symmetric solutions
is different from that of previous works~\cite{Crisostomi:2017lbg,%
Langlois:2017dyl,Dima:2017pwp,Crisostomi:2017pjs},
it is a good exercise to check here
that one can reproduce the previous result
in a nonrelativistic, weak-field limit.

We write
\begin{align}
\nu=\delta\nu,\quad \lambda = \delta \lambda,\quad X=-v^2+\delta X,
\end{align}
and derive linearized equations for a nonrelativistic source
with $P=0$.
It is straightfoward to derive
\begin{align}
{\cal F}_\lambda &\simeq 1+r\left(\delta\nu'+\frac{2f_X}{f} \delta X'\right),
\\
{\cal F}_1&\simeq -\frac{v^2 f}{2v^2f_X+fB_1},
\\
{\cal F}_2&\simeq \frac{f(\delta X-v^2\delta\nu)}{2v^2f_X+fB_1}
-\left(\frac{v^2}{v^2f_X+fB_1}\right)\frac{r^2\rho}{8},
\\
{\cal F}_3&\simeq \frac{\delta X-v^2\delta\nu}{rv^2}-4\pi G_Nr\rho
\notag \\ &\quad +2\pi G_N\left[-\frac{12v^2f_X}{f}
+\frac{(1-3B_1)B_1f}{v^2f_X+fB_1}
\right]r\rho
\notag \\ &\quad
+2\pi G_N \Upsilon_1 r^2\rho',
\end{align}
where we introduced
\begin{align}
8\pi G_N:=\left.\frac{1}{2f(1-3B_1)+4Xf_X}\right|_{X=-v^2},\label{def:GNinf}
\end{align}
and
\begin{align}
  \Upsilon_1:=\left.
  -\frac{(-2Xf_X+fB_1)^2}{f(-Xf_X +fB_1)}\right|_{X=-v^2}.\label{def:Up1}
\end{align}
We will see below that $G_N$ can indeed be regarded as
the Newton constant.
We then solve the following set of equations:
\begin{align}
\delta \lambda&={\cal F}_\lambda -1,\label{eq:Newton1}
\\
\delta X'&={\cal F}_1\delta\nu'+\frac{{\cal F}_2}{r},\label{eq:Newton2}
\\
\delta \nu'&={\cal F}_3.\label{eq:Newton3}
\end{align}
Combining Eqs.~(\ref{eq:Newton2}) and~(\ref{eq:Newton3}),
the following second-order equation for $\delta\nu$ can be derived,
\begin{align}
\delta\nu''+\frac{2}{r}\delta\nu'=
2G_N\left[
\frac{M'}{r^2}+\Upsilon_1\left(\frac{M''}{2r}+\frac{M'''}{4}\right)\right],
\label{eq:ddnu}
\end{align}
where $M(r)$ is the enclosed mass defined as
\begin{align}
M(r):=4\pi\int^r\rho(s) s^2\D s.
\end{align}
Equation~(\ref{eq:ddnu}) can be integrated to give
\begin{align}
\delta\nu'=\frac{C_0}{r^2}
+2G_N\left(\frac{M}{r^2}+\frac{\Upsilon_1}{4}M''\right),
\end{align}
where $C_0$ is an integration constant.
Combining Eqs.~(\ref{eq:Newton2}) and~(\ref{eq:Newton3}) again,
we obtain
\begin{align}
\delta X &=v^2\delta \nu+ \frac{v^2C_0}{r}+\frac{2v^2G_N M}{r}
\notag \\ &\quad +
v^2G_N\left[ 1+\frac{2v^2f_X}{f}-\frac{(1-B_1)fB_1}{2(v^2f_X+fB_1)}
\right]M'.
\end{align}
Finally, we use Eq.~(\ref{eq:Newton1}) to get
\begin{align}
\delta \lambda = \frac{C_0}{r}+2G_Nr\left(
\frac{M}{r^2}-\frac{5\Upsilon_2}{4}\frac{M'}{r}
+\Upsilon_3 M''
\right),
\end{align}
where
\begin{align}
\Upsilon_2&:=\left.\frac{8Xf_X}{5f}\right|_{X=-v^2},\label{def:Up2}
\\
\Upsilon_3&:=\left.-\frac{B_1}{4}\left(
\frac{-2Xf_X+fB_1}{-Xf_X +fB_1}\right)\right|_{X=-v^2}.\label{def:Up3}
\end{align}
Imposing regularity at the center, we take $C_0=0$.

We may set $M'=0$ and $M''=0$ outside the source,
and then we have
$\delta\nu= - \delta\lambda = -2G_NM/r$,
which coincides with the solution in general relativity
if $G_N$ is identified as the Newton constant.
Gravity is modified only inside the matter source, and
we have seen that
there are three parameters, $\Upsilon_{1,2,3}$, that characterize
the deviation from the standard result.
They are subject to
\begin{align}
2\Upsilon_1^2-5\Upsilon_1\Upsilon_2-32\Upsilon_3^2=0,
\end{align}
so that actually only two of them are independent.
Note that in the case of the GLPV theory, one has $B_1=0$,
and hence $\Upsilon_2=2\Upsilon_1/5$ and $\Upsilon_3=0$.

To see that the previous result is correctly reproduced,
we perform a small coordinate transformation
\begin{align}
\varrho=r\left[1+\frac{1}{2}\int^r\frac{\delta\lambda(r')}{r'}\D r'\right].
\end{align}
The metric then takes the form
\begin{align}
\D s^2=-(1+2\Phi)\D t^2 +
(1-2\Psi)\left(\D\varrho^2+\varrho^2\D\Omega^2\right),
\end{align}
where
\begin{align}
\Phi =\frac{\delta\nu}{2},
\quad
\Psi = \frac{1}{2}\int^r\frac{\delta\lambda(r')}{r'}\D r'.
\label{eq:wfpsiphi}
\end{align}
Thus, we can confirm that Eq.~(\ref{eq:wfpsiphi}) reproduces
the previous result found in the
literature~\cite{Crisostomi:2017lbg,Langlois:2017dyl,Dima:2017pwp,Crisostomi:2017pjs}.

Constraints on the $\Upsilon$ parameters have been obtained from
astrophysical observations in the case of
$\Upsilon_3=0$~\cite{Koyama:2015oma,%
Sakstein:2015zoa,Sakstein:2015aac,Jain:2015edg,%
Sakstein:2016ggl,Sakstein:2016lyj,Salzano:2017qac,Babichev:2016jom,Sakstein:2016oel}.
For example, the mass-radius relation of white dwarfs yields the constraint
$-0.18<\Upsilon_1<0.27$~\cite{Jain:2015edg}.
This is valid even in the case of $\Upsilon_3\neq 0$, because
the constraint comes only from $\Phi$ in such nonrelativistic systems.
To probe $\Psi$, one needs nonrelativistic systems and/or
observations based on propagation of light rays such as
gravitational lensing, and a tighter constraint has been
imposed on $\Upsilon_1$
combining the information on $\Psi$~\cite{Sakstein:2016ggl}.
However, this relies on the assumption that $\Upsilon_3=0$.
For this reason, it is important to study
relativistic stars in theories with $\Upsilon_3\neq 0$.

Another constraint can be obtained from the Hulse-Taylor pulsar,
which limits the difference between $G_N$ and the effective
gravitational coupling for gravitational waves,
$G_{\rm GW}=1/16\pi f(-v^2)$~\cite{Jimenez:2015bwa}.
The constraint reads~\cite{Dima:2017pwp}
\begin{align}
\frac{G_{\rm GW}}{G_N}-1 \left.=
\frac{2Xf_X}{f}-3B_1\right|_{X=-v^2}
<{\cal O}(10^{-3}).\label{Ggwconstraint}
\end{align}
Note, however, that this constraint is
based on the assumption that the scalar radiation
does not contribute to the energy loss, whose validity must be
ascertained in the Vainshtein-breaking theories.

\section{Boundary conditions}

\subsection{Boundary conditions at the center}\label{subsec:center}

To derive the boundary conditions at the center of a star,
we expand
\begin{align}
&\nu=\nu_c+\frac{\nu_2}{2}r^2+\cdots,
\quad
X=X_c\left(1+\frac{X_2}{2}r^2+\cdots\right),
\notag
\\
&\rho=\rho_c+\frac{\rho_2}{2}r^2+\cdots,
\quad
P=P_c+\frac{P_2}{2}r^2+\cdots,
\end{align}
where
\begin{align}
X_c:=-e^{-\nu_c}v^2.\label{eq:Xc}
\end{align}
In deriving the relation~(\ref{eq:Xc}) we used regularity
at the center, $\phi'(t,0)=\psi'(0)=0$.
We then expand ${\cal F}_\lambda$, ${\cal F}_1$, ${\cal F}_2$,
and ${\cal F}_3$ around $r=0$ to obtain
\begin{align}
&{\cal F}_\lambda\simeq
1+a_\lambda r^2,
\quad
{\cal F}_1\simeq a_1,
\notag \\
&
{\cal F}_2\simeq a_2 r^2,\quad {\cal F}_3\simeq a_3r,
\end{align}
where
\begin{align}
a_\lambda&:=\nu_2+\frac{4X_cX_2f_X(X_c)-P_c}{2f(X_c)},
\\
a_1&:=\frac{X_cf(X_c)}{f(X_c)B_1(X_c)-2X_cf_X(X_c)},
\end{align}
while $a_2$ and
$a_3$ are similar but slightly more messy.

Equation~(\ref{eq:elambda}) implies
$e^\lambda\simeq 1+a_\lambda r^2$,
so that we find
\begin{align}
\lambda=a_\lambda r^2+\cdots.
\end{align}
Equations~(\ref{eq:dX}) and~(\ref{eq:dnu}) reduce to
the following algebraic equations
\begin{align}
X_cX_2=a_1 \nu_2+a_2,\quad \nu_2=a_3,
\end{align}
leading to
\begin{align}
\nu_2&=\frac{8\pi G_c}{3}\left(\rho_c+3P_c\right)
\notag \\
&\quad +4\pi G_c\left[\eta_1 \rho_c+\left(5\eta_2+12\eta_3\right)
P_c\right],
\\
X_2&=-8\pi G_c
\left[
2\rho_c+3P_c-\frac{4\eta_3}{\eta_1+4\eta_3}(\rho_c-3P_c)
\right],
\end{align}
where
\begin{align}
\eta_1&:=\left.-\frac{(-2Xf_X+fB_1)^2}{f(-Xf_X+fB_1)}\right|_{X=X_c},
\\
\eta_2&:=\left.\frac{8Xf_X}{5f}\right|_{X=X_c},
\\
\eta_3&:=\left.-\frac{B_1}{4}\left(
\frac{-2Xf_X+fB_1}{-Xf_X +fB_1}\right)\right|_{X=X_c},
\end{align}
and we defined the effective gravitational constant
at the center as
\begin{align}
8\pi G_c:=\left.\frac{1}{2f(1-3B_1)+4Xf_X}\right|_{X=X_c}.
\end{align}

The above quantities are defined following
Eqs.~(\ref{def:GNinf}), (\ref{def:Up1}), (\ref{def:Up2}), and~(\ref{def:Up3}),
but now they are evaluated at the center, $X=X_c$.
If gravity is sufficiently weak and the Newtonian approximation is valid,
we have $|\nu_c|\ll 1$ and hence $X_c\simeq -v^2$, leading to
$G_c\simeq G_N$ and $\eta_i\simeq \Upsilon_i$.
In this case, corrections to the standard expression for the metric
near $r=0$ are small as long as $\Upsilon_i\ll 1$.
However, if the system is in the strong gravity regime,
$X_c$ may differ significantly from $-v^2$ and therefore
the internal structure of relativistic stars may be modified
notably even in theories where only a tiny correction is expected
in the Newtonian regime.

Finally,
from Eq.~(\ref{eq:fluid}) we get
\begin{align}
P_2=-\frac{\nu_2}{2}(\rho_c+P_c).
\end{align}

Now,
given $\nu_c$ and $\rho_c$ (or $P_c$),
Eqs.~(\ref{eq:fluid}), (\ref{eq:dX}), and~(\ref{eq:dnu})
can be integrated from the center outward.
Let us move to the boundary conditions outside the star.

\subsection{Exterior solution}

For $\rho=P=0$, Eqs.~(\ref{eq:dX}) and~(\ref{eq:dnu})
reduce to the following simple set of equations:
\begin{align}
X'=0,\quad \nu'=-\frac{X+e^{-\nu}v^2}{rX}.
\end{align}
This can be integrated to give
\begin{align}
X=-v^2,\quad e^{\nu}=1-\frac{{\cal C}}{r},
\end{align}
where ${\cal C}$ is an integration constant and
we imposed that
$\psi'\to 0$ as $r\to \infty$.
We then have
\begin{align}
e^\lambda &= {\cal F}_\lambda(\nu,\nu',-v^2,0,0)=1+\nu' r
\notag \\
&=\left(1-\frac{{\cal C}}{r}
\right)^{-1}.
\end{align}
The exterior metric is thus obtained exactly
without linearizing the equations,
which coincides with the Schwarzschild solution in general relativity.
The stellar interior must be matched to this exterior solution.
It will be convenient to write
\begin{align}
{\cal C}=2G_N\mu,
\end{align}
because then $\mu$ is regarded as the mass of the star.

\subsection{Matching at the stellar surface}
\label{subsec:match}

The stellar surface, $r=R$, is defined by $P(R)=0$.
The induced metric is required to be continuous across the surface,
so that $\nu$ must be continuous there.
Since $X=\phi_\mu\phi^\mu$ is a spacetime scalar,
it is reasonable to assume that
this quantity is also continuous across the surface.
We thus have the two conditions imposed at $r=R$:
\begin{align}
e^{\nu(R)}&= 1-\frac{2G_N\mu}{R},\label{jc:2}
\\
X(R)&=-v^2.\label{jc:1}
\end{align}
We tune the central value $\nu_c$ in order for the solution
to satisfy Eq.~(\ref{jc:1}). The second condition~(\ref{jc:2})
is used to determine the integration constant $\mu$.
As we have seen in the previous section,
$\mu=M(R)$ in the nonrelativistic, weak-field limit.
In the present case, however, $\mu$ does not necessarily
coincide with $M(R)$ because the nonrelativistic and weak-field
approximations are not justified
in general for our interior solutions.

Note that in general we may have $\rho_-':=\rho'(R_-)\neq 0$
while $\rho'(R_+)=0$, where
$R_\pm:=\lim_{\varepsilon\to 0} R(1\pm\varepsilon)$.
As a particular feature of the DHOST theories
with partial breaking of the Vainshtein mechanism,
the right-hand side of Eq.~(\ref{eq:dnu}) depends on
$\rho'$~\cite{Babichev:2016jom}. This implies that $\nu'$ is discontinuous
across the stellar surface.
Then, from Eq.~(\ref{eq:dX}) we see that
$X'$ is also discontinuous across the surface.
Furthermore, since the right hand side of Eq.~(\ref{eq:elambda})
depends on $\nu'$ and $X'$,
$\lambda$ is also discontinuous there in general.
With some manipulation we see that
\begin{align}
&1-e^{[\lambda(R_-)-\lambda(R_+)]/2}
\notag \\
&=\pi G_N\rho_-'R^3 B_1\left.\left[2e^{-\nu(R)}-\frac{B_1f}{B_1f-Xf_X}\right]
\right|_{X=-v^2},
\end{align}
which shows that $\lambda$ is nevertheless continuous in theories with $B_1=0$.
However, it is found that
\begin{align}
&X'(R_+)-X'(R_-)
\notag \\ &
=
2\pi G_N\rho_-'R^2X
\left.\left[2e^{-\nu(R)}-\frac{B_1f}{B_1f-Xf_X}\right]
\right|_{X=-v^2},
\end{align}
and therefore $X'$ is discontinuous even if $B_1=0$.
This is also the case for $\nu'$.
In the next section,
we will show our numerical results in which one can find these
discontinuities.

\section{Numerical results}

As a specific example, we study the model of Ref.~\cite{Crisostomi:2017pjs}:
\begin{align}
f=\frac{\mpl^2}{2}+\alpha X^2,\quad A_3=-8\alpha-\beta,
\end{align}
where $\alpha$ and $\beta$ are constants.
Note that we are using the notation such that $8\pi G_N\neq \mpl^{-2}$.
We have
\begin{align}
B_1=-\frac{\beta X^2}{2(\mpl^2+2\alpha X^2)},
\end{align}
and hence
the model with $\beta\neq 0$ is more general than the GLPV theory.
For this choice of the functions the degeneracy conditions~(\ref{eq:degn1})
and~(\ref{eq:degn2}) leads to
\begin{align}
A_4&=\frac{\mpl^2(8\alpha+\beta)+(16\alpha^2-6\alpha\beta-\beta^2/4)X^2}{\mpl^2+2\alpha X^2},
\\
A_5&=\frac{\beta(8\alpha+\beta)X}{\mpl^2+2\alpha X^2}.
\end{align}
This model, with the addition of the lower order Horndeski terms,
admits viable self-accelerating cosmological solutions~\cite{Crisostomi:2017pjs},
and therefore is interesting.

Hereafter we will use the dimensionless parameters
defined as
\begin{align}
\overline{\alpha}:=\frac{\alpha v^4}{\mpl^2},\quad
\overline{\beta}:=\frac{\beta v^4}{\mpl^2}.
\end{align}
The parameters
that characterize Vainshtein breaking in the Newtonian regime,
$\Upsilon_i$, can then be estimated as
\begin{align}
\Upsilon_i \sim \overline{\alpha},\overline{\beta}.
\end{align}
From Eq.~(\ref{Ggwconstraint}) we also estimate
\begin{align}
\frac{G_{\rm GW}}{G_N}-1\sim \overline{\alpha},\overline{\beta}.
\end{align}
Therefore, by taking sufficiently small $\overline{\alpha}$ and
$\overline{\beta}$ (say, ${\cal O}(10^{-3})$),
current constraints can be evaded.
In the following numerical calculations,
we will employ such small values of the parameters.

The equation of state we use is given by
\begin{align}
\rho=\left(\frac{P}{K}\right)^{1/2}+P,
\end{align}
with $K=7.73\times 10^{-3}\, (8\pi G_N)^3\msun^2$
($K=123\,\msun^2$ in the units where $G_N=1$),
which has been used frequently in the modifed gravity
literature~\cite{Cisterna:2015yla,Cisterna:2016vdx,Maselli:2016gxk,Silva:2016smx,%
Babichev:2016jom}.
With this simple equation of state we focus on
the qualitative nature of the solutions.

\begin{figure}[tbp]
  \begin{center}
  \includegraphics[width=80mm]{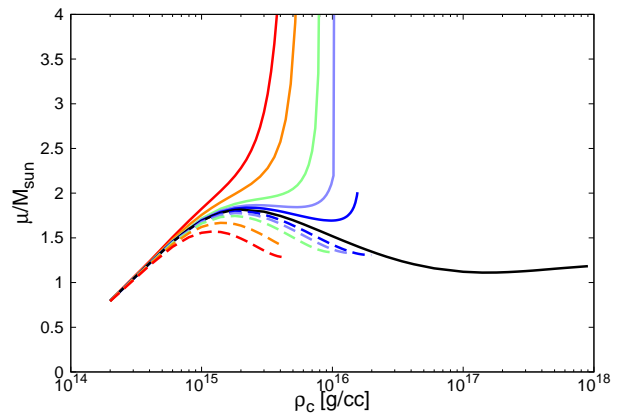}
  \end{center}
  \caption{The mass ($\mu$) versus central density
  diagram for $\overline{\beta}=0$.
  The parameters are given by
  $\overline{\alpha} = 2\times 10^{-3}, 10^{-3}, 4\times 10^{-4},%
  2\times 10^{-4}, 10^{-4}, 0$ (GR) (solid lines, from top to bottom), and
  $\overline{\alpha}=-10^{-4}, -2\times 10^{-4}, -4\times 10^{-4}, -10^{-3}, -2\times 10^{-3}$
  (dashed lines, from top to bottom).}
  \label{fig:plot_valpha_rhoM.eps}
\end{figure}

\begin{figure}[tbp]
  \begin{center}
  \includegraphics[width=80mm]{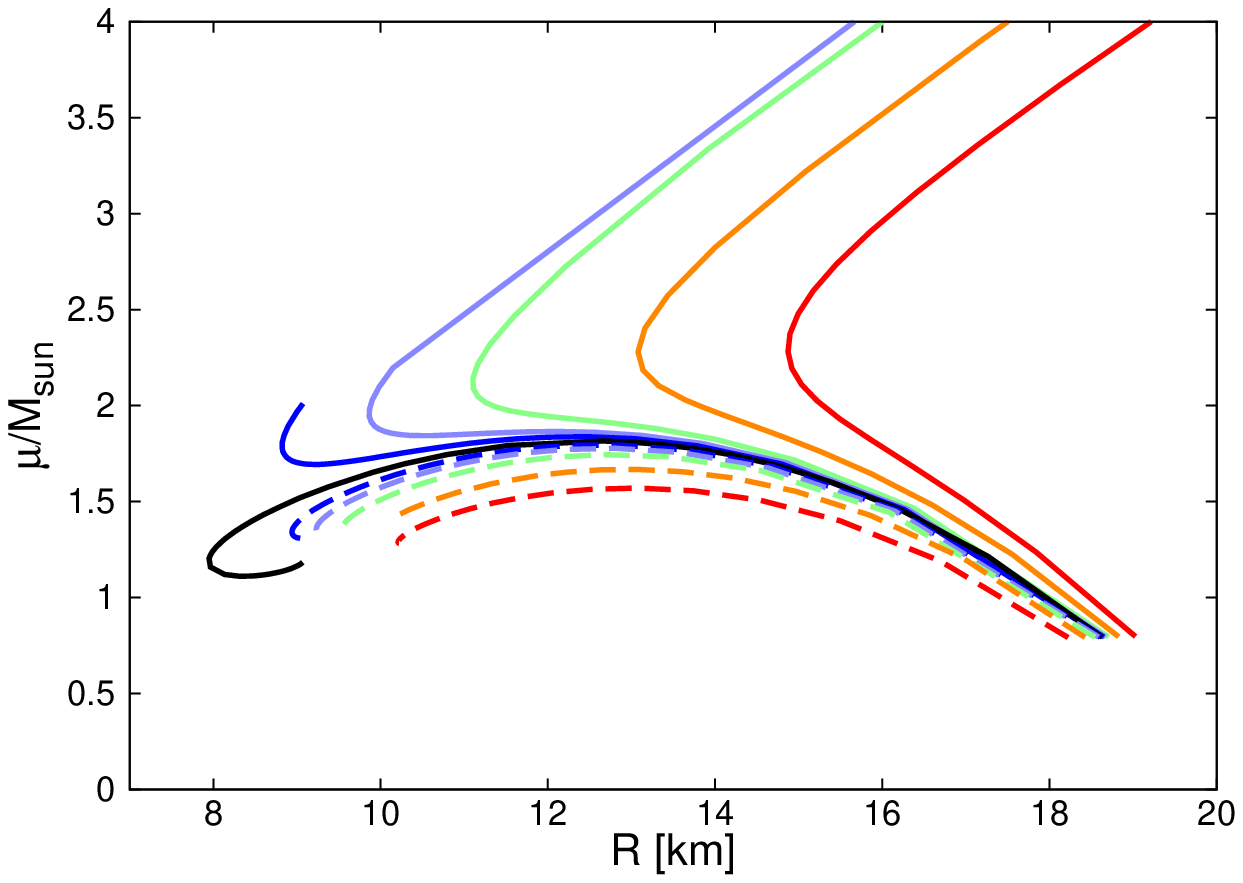}
  \end{center}
  \caption{The mass ($\mu$) versus radius diagram for $\overline{\beta}=0$.
  The parameters are given by
  $\overline{\alpha} = 2\times 10^{-3}, 10^{-3}, 4\times 10^{-4},%
  2\times 10^{-4}, 10^{-4}, 0$ (GR) (solid lines, from right to left), and
  $\overline{\alpha}=-10^{-4}, -2\times 10^{-4}, -4\times 10^{-4}, -10^{-3}, -2\times 10^{-3}$
  (dashed lines, from top to bottom).}%
  \label{fig:plot_valpha_RM.eps}
\end{figure}

\begin{figure}[tbp]
  \begin{center}
  \includegraphics[width=80mm]{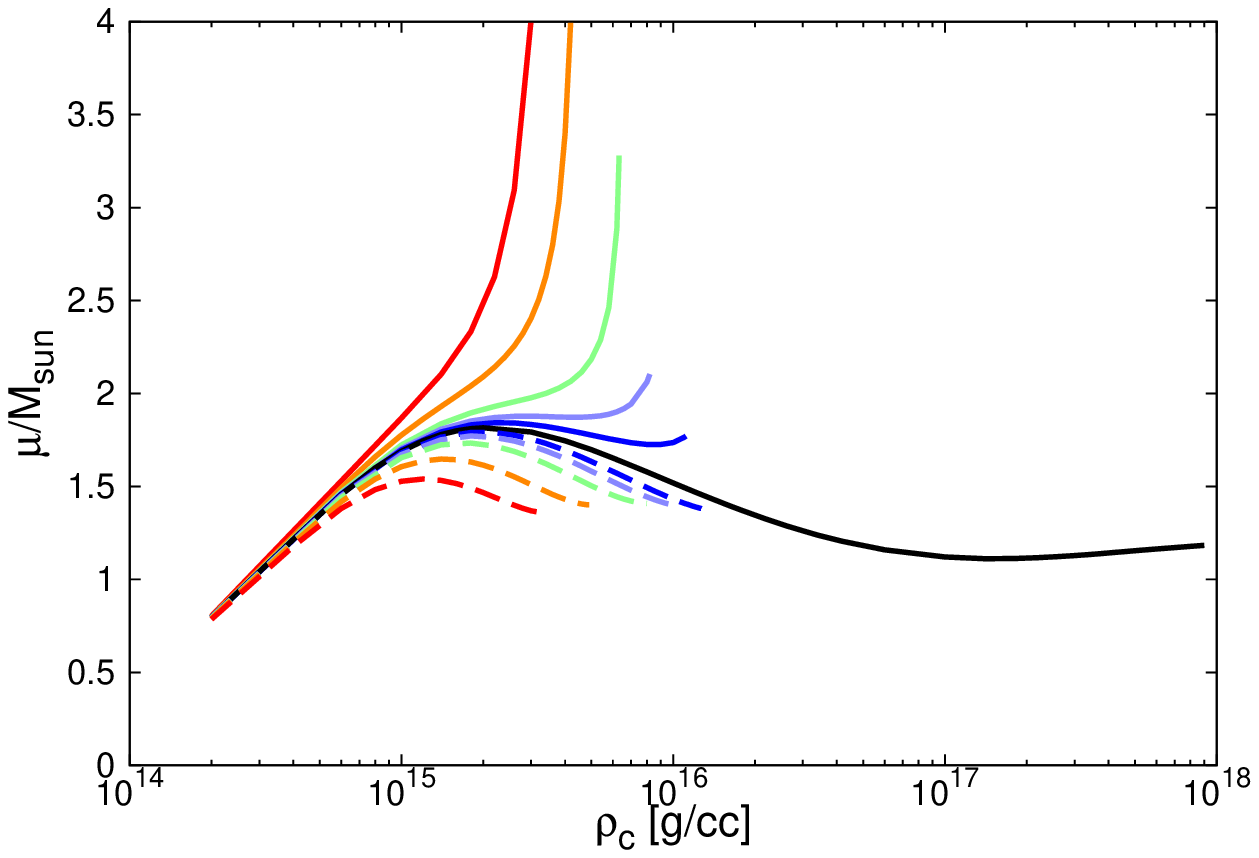}
  \end{center}
  \caption{The mass ($\mu$) versus central density
  diagram for $\overline{\alpha}=0$.
  The parameters are given by
  $\overline{\beta} = 2\times 10^{-3}, 10^{-3}, 4\times 10^{-4},%
  2\times 10^{-4}, 10^{-4}, 0$ (GR) (solid lines, from top to bottom), and
  $\overline{\beta}=-10^{-4}, -2\times 10^{-4}, -4\times 10^{-4}, -10^{-3}, -2\times 10^{-3}$
  (dashed lines, from top to bottom).}
  \label{fig:plot_vbeta_rhoM.eps}
\end{figure}

\begin{figure}[tbp]
  \begin{center}
  \includegraphics[width=80mm]{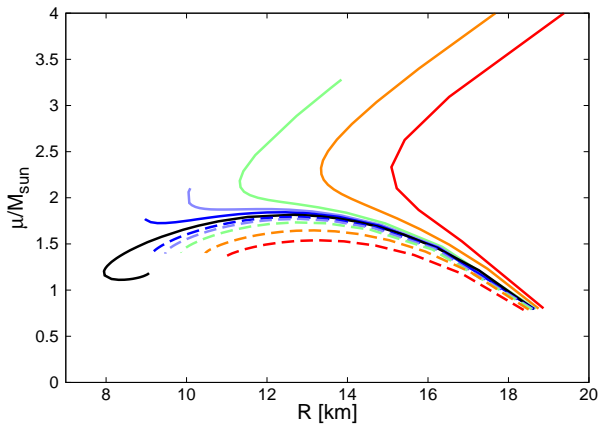}
  \end{center}
  \caption{The mass ($\mu$) versus radius
  diagram for $\overline{\alpha}=0$.
  The parameters are given by
  $\overline{\beta} = 2\times 10^{-3}, 10^{-3}, 4\times 10^{-4},%
  2\times 10^{-4}, 10^{-4}, 0$ (GR) (solid lines, from right to left), and
  $\overline{\beta}=-10^{-4}, -2\times 10^{-4}, -4\times 10^{-4}, -10^{-3}, -2\times 10^{-3}$
  (dashed lines, from top to bottom).}%
  \label{fig:plot_vbeta_RM.eps}
\end{figure}

We start with the theories with $\overline{\beta}=0$
and focus on the effect of $\overline{\alpha}$.
Figures~\ref{fig:plot_valpha_rhoM.eps}
and~\ref{fig:plot_valpha_RM.eps}
show the mass ($\mu$) versus central density relation and
the mass versus radius relation, respectively.
In all cases $\overline{\alpha}$ is taken to be very small so that
the Vainshtein-breaking effect is not significant in the Newtonian regime.
It can be seen that
for fixed $\rho_c$ or $R$ the mass is larger (smaller)
for $\overline{\alpha}>0$ ($\overline{\alpha}<0$)
than in the case of general relativity (GR).
Interestingly, there is a maximum central density, $\rho_{c,{\rm max}}$,
above which no solution can be found.
This property is similar to what was found
in Ref.~\cite{Cisterna:2015yla},
where the subclass of the Horndeski theory
having derivative coupling to the Einstein tensor was studied.
For $\overline{\alpha}\lesssim 2\times 10^{-4}$,
we see that $\mu<\infty$ as $\rho_c\to\rho_{c,{\rm max}}$,
but for $\overline{\alpha}\gtrsim 2\times 10^{-4}$
we find that $\mu\to \infty$ and $R\to\infty$
as $\rho_c\to\rho_{c,{\rm max}}$.
Therefore, in the latter case there are solutions at high densities
that are very different from relativistic stars in GR.
Note that this occurs even for a tiny value of $\overline{\alpha}$.

\begin{figure}[tbp]
  \begin{center}
  \includegraphics[width=80mm]{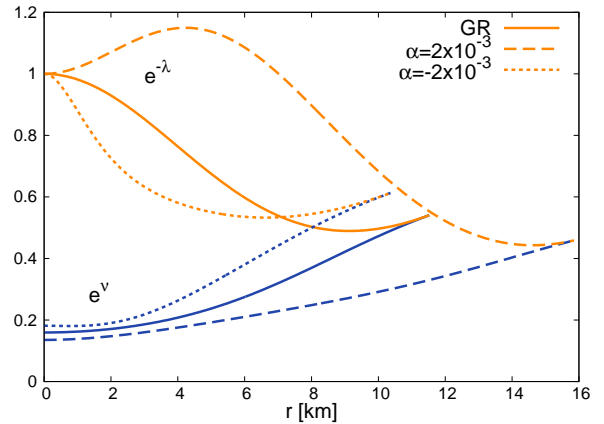}
  \end{center}
  \caption{Metric components $e^{\nu}$ and $e^{-\lambda}$ as a function of $r$.
  The parameters are given by $\overline{\alpha}=2\times 10^{-3}, \overline{\beta}=0$
  (dashed lines)
  and $\overline{\alpha}=-2\times 10^{-3}, \overline{\beta}=0$
  (dotted lines). Solid lines represent the result of GR.}%
  \label{fig:plot_valpha_metric.eps}
\end{figure}

\begin{figure}[tbp]
  \begin{center}
  \includegraphics[width=80mm]{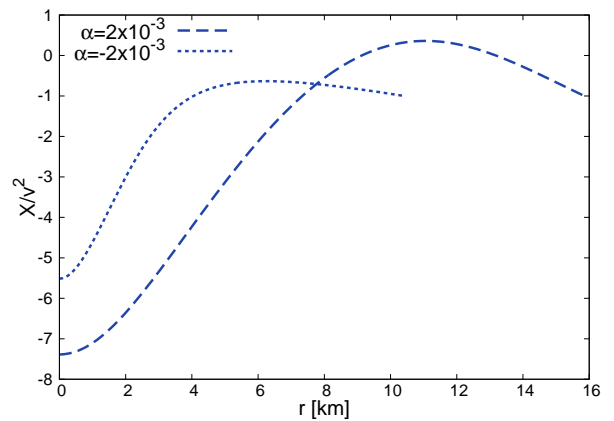}
  \end{center}
  \caption{$X/v^2$ as a function of $r$.
  The parameters are given by $\overline{\alpha}=2\times 10^{-3}, \overline{\beta}=0$
  (dashed line)
  and $\overline{\alpha}=-2\times 10^{-3}, \overline{\beta}=0$
  (dotted line).}%
  \label{fig:plot_valpha_x.eps}
\end{figure}

\begin{figure}[tbp]
  \begin{center}
  \includegraphics[width=80mm]{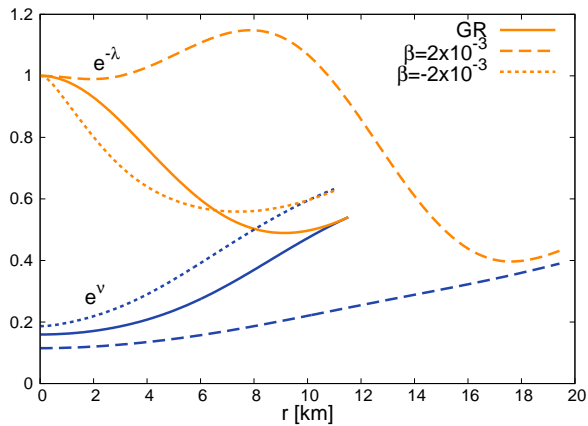}
  \end{center}
  \caption{Metric components $e^{\nu}$ and $e^{-\lambda}$ as a function of $r$.
  The parameters are given by $\overline{\alpha}=0, \overline{\beta}=2\times 10^{-2}$
  (dashed lines)
  and $\overline{\alpha}=0, \overline{\beta}=-2\times 10^{-2}$
  (dotted lines). Solid lines represent the result of GR.}%
  \label{fig:plot_vbeta_metric.eps}
\end{figure}

\begin{figure}[tbp]
  \begin{center}
  \includegraphics[width=80mm]{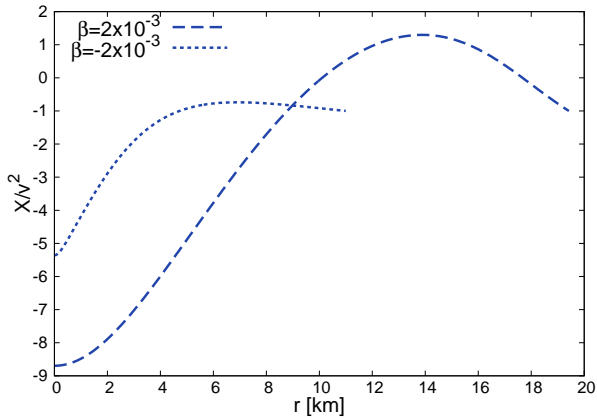}
  \end{center}
  \caption{$X/v^2$ as a function of $r$.
  The parameters are given by $\overline{\alpha}=0, \overline{\beta}=2\times 10^{-2}$
  (dashed line)
  and $\overline{\alpha}=0, \overline{\beta}=-2\times 10^{-2}$
  (dotted line).}%
  \label{fig:plot_vbeta_x}
\end{figure}

Next, we fix $\overline{\alpha}=0$ and
draw the same diagrams for different (small)
values of $\overline{\beta}$.
The results are presented in Figs.~\ref{fig:plot_vbeta_rhoM.eps}
and~\ref{fig:plot_vbeta_RM.eps}, which are
seen to be qualitatively similar to
Figs.~\ref{fig:plot_valpha_rhoM.eps}
and~\ref{fig:plot_valpha_RM.eps}, respectively.
Therefore, although $\overline{\beta}$ is supposed to
signal the ``beyond GLPV'' effects, they are not manifest
and the roles of the two parameters
$\overline{\alpha}$ and $\overline{\beta}$
in relativistic stars
are qualitatively similar.
We have performed numerical calculations
for more general cases with
$\overline{\alpha}\neq 0$ and $\overline{\beta}\neq 0$,
and confirmed that they also
lead to qualitatively similar results.

Just for reference
some examples of radial profiles of
the metric and $X$ in the stellar interior are presented in
Figs.~\ref{fig:plot_valpha_metric.eps} and~\ref{fig:plot_valpha_x.eps}
for $(\overline{\alpha},\overline{\beta})=(\pm 2\times 10^{-3},0)$
and in Figs.~\ref{fig:plot_vbeta_metric.eps} and~\ref{fig:plot_vbeta_x}
for $(\overline{\alpha},\overline{\beta})=(0,\pm 2\times 10^{-3})$.
As mentioned in Sec.~\ref{subsec:match},
one can find that $X'$ is
nonvanishing at the surface, leading to the discontinuity of $X'$, and
that $e^{-\lambda}$ does not agree with $e^{\nu}$ there
in Fig.~\ref{fig:plot_vbeta_metric.eps}, indicating the discontinuity
of $e^{-\lambda}$ since $e^{\nu}$ must be continuous.

\section{Conclusions}

In this paper, we have studied the Tolman-Oppenheimer-Volkoff system in
degenerate higher-order scalar-tensor (DHOST) theory
that is consistent with the GW170817 constraint on
the speed of gravitational waves.
Although the field equations are apparently of higher order,
we have reduced them to a first-order system
by combining the different components.
This is possible because the theory we are considering is degenerate.

In DHOST theories that are more general than Horndeski,
breaking of the Vainshtein screening mechanism occurs
inside matter~\cite{Kobayashi:2014ida,Crisostomi:2017lbg,Langlois:2017dyl,Dima:2017pwp},
which would modify the interior structure of stars.
Assuming a simple concrete model of DHOST theory with two parameters
and the equation of state, we have solved numerically the field equations.
The parameters were chosen so that the Vainshtein-breaking effect
in the Newtonian regime is suppressed by
the factor $\Upsilon_i\lesssim 10^{-3}$.
Nevertheless, we have found a possible large modification
in the mass-radius relation. This is significant in particular
at densities as high as the maximum above which no solutions
can be obtained.

In this paper, we have focused on the rather qualitative nature
of relativistic stars in the DHOST theory, but
it would be important to employ more realistic equations of state
for testing the theory against astrophysical observations.
This is left for further study.
It would be also interesting to explore to what extent
the modification to the stellar structure depends on
the concrete form of the DHOST Lagrangian.
We hope to come back to this question in a future study.

\acknowledgments

This work was supported in part by
MEXT KAKENHI Grant Nos.~JP15H05888, JP16H01102 and JP17H06359 (T.K.);
JSPS KAKENHI Grant Nos.~JP16K17695 (T.H.) and JP16K17707 (T.K);
and MEXT-Supported Program for the Strategic Research Foundation at Private Universities,
2014-2018 (S1411024) (T.H. and T.K.).
\clearpage

\begin{widetext}
  \appendix

\section{Explicit form of ${\cal F}_1,{\cal F}_2$, and ${\cal F}_3$}
\label{App:f1f2f3}

Here we present the explicit expression for ${\cal F}_1$,
${\cal F}_2$, and ${\cal F}_3$ that appear in Eqs.~(\ref{eq:dX}) and (\ref{eq:dnu}):
  \begin{align}
  {\cal F}_1=\frac{U_1}{V},\quad {\cal F}_2=\frac{U_2}{V},
  \quad
  {\cal F}_3=\frac{U_3}{W},
  \end{align}
  where
  \begin{footnotesize}
  \begin{align}
  \frac{U_1}{2 f e^{\nu } X^2}
  &= 2 r^2 v^2 (P+\rho ) \left(B_1
   f-X f_X\right)+e^{\nu } X \left\{
   B_1 f \left[r^2
   (\rho -5 P)-8 f\right]+2 X f_X \left(4 f+P
   r^2\right)\right\},
  \\
  \frac{U_2}{2 f e^{\nu } X^2}
  &= -2 v^2 \left(B_1 f-X
   f_X\right) \left[4 f+r^2 (P-\rho )\right]
+e^{\nu } X \left\{B_1 f
 \left[r^2 (\rho -5 P)-8 f\right]+2 X f_X \left(4
 f+P r^2\right)\right\},
  \\
  V&=-8 r^2 v^4 (P+\rho ) \left(B_1 f-X f_X\right){}^2+2
   e^{\nu } v^2 X \left(X f_X-B_1 f\right) \left\{B_1 f
   \left[r^2 (5 \rho -9 P)-20 f\right]+4 X f_X \left(4
   f+P r^2\right)\right\}
\notag \\ &\quad
   +3 B_1 f e^{2 \nu } X^2
   \left\{B_1 f \left[8 f+r^2 (5 P-\rho )\right]-2 X
   f_X \left(4 f+P r^2\right)\right\},
\\
    U_3&=
    -16 e^{3 \nu } f r^4 f_X^2 \bigl[
    B_1 (X f_X-2
    f)+X (f B_{1X}+f_X-X
    f_{XX})\bigr]
     (\rho -3
    P)^2 X^5
    \notag \\ &\quad
    +8 e^{2 \nu } r^2 (f B_1-X
    f_X){}^2
    \bigl\{
    16 \bigl[e^{\nu }
    f_X+(v^2+e^{\nu } X)
    f_{XX}\bigr] (3 P-\rho ) f^2
    \notag \\ &\quad
    +2
    \bigl[
    f_{XX} (\rho -3 P)
    ((2 v^2+e^{\nu } X) \rho -(2
    v^2+5 e^{\nu } X) P) r^2+e^{\nu } f_X
    (\rho ^2-8 P \rho +15
    P^2) r^2
    \notag \\ &\quad
    +4 f_X^2 (6
    (v^2+2 e^{\nu } X) \rho -18 (v^2+2
    e^{\nu } X) P+r (v^2+3 e^{\nu }
    X) \rho ')
    \bigr] f
    \notag \\ &\quad
    +r^2 f_X^2
    \bigl[(
    -408 e^{\nu } X P^2+(8
    (21 v^2+29 e^{\nu } X) \rho +r (28
    v^2+33 e^{\nu } X) \rho ') P
    \notag \\ &\quad
    -\rho
    (8 (7 v^2+4 e^{\nu } X) \rho +r
    (16 v^2+9 e^{\nu } X) \rho
    ')
    \bigr]
    \bigr\} X^4
    \notag \\ &\quad
    -2 e^{2 \nu } r^2 f_X
    (4 f B_1-4 X f_X) (3 P-\rho
    )
    \bigl\{
    -e^{\nu } r^2 f_X^2 (10 \rho -30
    P+3 r \rho ') X^3
    \notag \\ &\quad
    +16 f^2 (v^2+e^{\nu }
    X) \bigl[B_1 (X f_X-2 f)+X (f
    B_{1X}+f_X-X f_{XX})\bigr]
    \notag \\ &\quad
    +2 f r^2
    \bigl[
    \rho  (e^{\nu } X^3
    f_{XX}-(2 v^2+e^{\nu } X)
    (B_1 (X f_X-2 f)
    +X (f
    B_{1X}+f_X-X
    f_{XX})))
    \notag \\ &\quad
    -(3 e^{\nu }
    X^3 f_{XX}-(2 v^2+5 e^{\nu } X)
    (B_1 (X f_X-2 f)+X (f
    B_{1X}+f_X-X f_{XX})))
    P
    \bigr]
    \bigr\} X^3
    \notag \\ &\quad
    +8 e^{\nu } (f B_1-X
    f_X){}^3
    \bigl\{
    f_X \bigl[
    2 (68 v^4-24
    e^{\nu } X v^2-305 e^{2 \nu } X^2)
    P^2+(4 (8 v^4+122 e^{\nu }
    X v^2+73 e^{2 \nu } X^2) \rho
    \notag \\ &\quad
    +r (12
    v^4+76 e^{\nu } X v^2+39 e^{2 \nu } X^2) \rho
    ') P-(2 v^2+e^{\nu } X) \rho
    (52 \rho  v^2+34 e^{\nu } X \rho +r (14
    v^2+9 e^{\nu } X) \rho ')
    \bigr] r^4
    \notag \\ &\quad
    +2
    f \bigl[
    e^{\nu } (\rho -5 P)
    ((2 v^2+e^{\nu } X) \rho -(2
    v^2+5 e^{\nu } X) P) r^2+8 f_X (3
    (2 v^4+9 e^{\nu } X v^2+5 e^{2 \nu }
    X^2) \rho
    \notag \\ &\quad
     +3 (2 v^4-9 e^{\nu } X v^2-19
    e^{2 \nu } X^2) P+r (v^4+6 e^{\nu } X
    v^2+3 e^{2 \nu } X^2) \rho ')
    \bigr]
    r^2+128 e^{\nu } f^3 (v^2+e^{\nu } X)
    \notag \\ &\quad
    +16
    e^{\nu } f^2 \bigl[-(3 v^2+2 e^{\nu } X)
    \rho  r^2+(7 v^2+10 e^{\nu } X) P r^2-8
    X (v^2+e^{\nu } X) f_X\bigr]
    \bigr\}
    X^2
    \notag \\ &\quad
    -8 (f B_1-X f_X){}^4
    \bigl\{
    4 (16
    v^6-60 e^{\nu } X v^4+20 e^{2 \nu } X^2 v^2+75 e^{3
    \nu } X^3) P^2 r^4
    \notag \\ &\quad
    +(2
    v^2+e^{\nu } X)^2 (4 v^2+3 e^{\nu }
    X) \rho  (4 \rho +r \rho ') r^4-8
    e^{\nu } f X (2 v^2+e^{\nu } X) \bigl[6
    (4 v^2+3 e^{\nu } X) \rho +r (5
    v^2+3 e^{\nu } X) \rho '\bigr] r^2
    \notag \\ &\quad
    -P
    \bigl[-(2 v^2+e^{\nu } X) (8
    (8 v^4-12 e^{\nu } X v^2-15 e^{2 \nu }
    X^2) \rho +r (8 v^4-18 e^{\nu } X v^2-15
    e^{2 \nu } X^2) \rho ') r^2
    \notag \\ &\quad
    -48 e^{\nu }
    f X (-8 v^4+10 e^{\nu } X v^2+15 e^{2 \nu }
    X^2)
    \bigr] r^2+384 e^{2 \nu } f^2 X^2
    (v^2+e^{\nu } X)
    \bigr\},
    \\
   \frac{W}{r}&=16 e^{3 \nu } f r^4 f_X^2 \bigl[B_1 (X
   f_X-2 f)+X (f B_{1X}+f_X-X
   f_{XX})\bigr] (\rho -3
   P)^2 X^5
   \notag \\ &\quad
   -4 e^{2 \nu } r^2 (f B_1-X
   f_X){}^2
   \bigl\{
   \bigl[3 r^2 (16 v^2-77
   e^{\nu } X) P^2+4 \bigl((3
   v^2+71 e^{\nu } X) \rho  r^2+6 f (v^2-9
   e^{\nu } X)\bigr) P
   \notag \\ &\quad
   +\rho  (24 f
   (v^2+7 e^{\nu } X)-r^2 (36 v^2+61
   e^{\nu } X) \rho )\bigr] f_X^2-4
   e^{\nu } f (\rho -3 P) (-\rho
   r^2+5 P r^2+8 f) f_X
   \notag \\ &\quad
   -4 f f_{XX}
   (\rho -3 P) \bigl[-2 v^2 \rho
   r^2-(2 v^2-5 e^{\nu } X) P r^2+e^{\nu }
   X (8 f-r^2 \rho )\bigr]
   \bigr\} X^4
   \notag \\ &\quad
   +e^{2
   \nu } r^2 f_X (4 f B_1-4 X f_X) (3
   P-\rho ) \bigl\{
   -e^{\nu } r^2 f_X^2 (17
   \rho -15 P) X^3+32 e^{\nu } f^2 \bigl[B_1
   (X f_X-2 f)+X (f B_{1X}+f_X-X
   f_{XX})\bigr] X
   \notag \\ &\quad
   +4 f r^2 \bigl[\rho
   \bigl(e^{\nu } X^3 f_{XX}-(2 v^2+e^{\nu
   } X) (B_1 (X f_X-2 f)+X
   (f B_{1X}+f_X-X
   f_{XX}))\bigr)
   \notag \\ &\quad
   -\bigl( 3 e^{\nu }
   f_{XX} X^3+(2 v^2-5 e^{\nu } X)
   (B_1 (X f_X-2 f)+X (f
   B_{1X}+f_X-X f_{XX}))\bigr)
   P\bigr]
   \bigr\} X^3
   \notag \\ &\quad
   -4 e^{\nu } (f B_1-X
   f_X){}^3 \bigl\{4 e^{\nu } f (-\rho
   r^2+5 P r^2+8 f) \bigl[-2 v^2 \rho
   r^2-(2 v^2-5 e^{\nu } X) P r^2+e^{\nu }
   X (8 f-r^2 \rho )\bigr]
   \notag \\ &\quad
   -f_X \bigl[5
   (20 v^4-40 e^{\nu } X v^2+79 e^{2 \nu }
   X^2) P^2 r^4+4 v^4 \rho
   (12 f-5 r^2 \rho ) r^2+36 e^{\nu } v^2 X
   \rho  (3 r^2 \rho -4 f) r^2
   \notag \\ &\quad
   +2
   \bigl((40 v^4-46 e^{\nu } X v^2-199 e^{2 \nu }
   X^2) \rho  r^2+24 f (v^4-3 e^{\nu } X
   v^2+18 e^{2 \nu } X^2)\bigr) P r^2
   +e^{2 \nu
   } X^2 (71 \rho ^2 r^4-480 f \rho  r^2+256
   f^2)\bigr]\bigr\} X^2
   \notag \\ &\quad
   +4 (f B_1-X
   f_X){}^4 \bigl\{-48 r^4 \rho ^2 v^6+12 e^{\nu
   } r^2 X \rho  (\rho  r^2+20 f) v^4+8
   e^{2 \nu } r^2 X^2 \rho  (9 r^2 \rho -47
   f) v^2
   \notag \\ &\quad
   -r^4 (48 v^6-180 e^{\nu } X
   v^4+200 e^{2 \nu } X^2 v^2-225 e^{3 \nu }
   X^3) P^2+3 e^{3 \nu } X^3
   (9 \rho ^2 r^4-104 f \rho  r^2+256
   f^2)
   \notag \\ &\quad
   -4 r^2 \bigl[24 r^2 \rho  v^6-12 e^{\nu }
   X (4 \rho  r^2+5 f) v^4+2 e^{2 \nu } X^2
   (16 \rho  r^2+47 f) v^2-15 e^{3 \nu }
   X^3 (14 f-3 r^2 \rho )\bigr]
   P\bigr\}.
  \end{align}
\end{footnotesize}
\end{widetext}




\end{document}